\documentclass[fleqn,twoside]{article}
\begin{document}
\begin{center}
\begin{large}
ON THE ISING-HEISENBERG MODEL WITH THE DOUBLY DECORATED
       NETWORK STRUCUTRE I
\end{large}
\newline
\newline
\begin{large}
\hspace*{1cm} Jozef Stre\v{c}ka and Michal Ja\v{s}\v{c}ur
\newline
\end{large}
Department of Theoretical Physics and Astrophysics, Faculty of Science, \\
P. J. \v{S}af\'{a}rik University, Moyzesova 16, 041  54 Ko\v{s}ice,
Slovak Republic, E-mail:jozkos@pobox.sk, jascur@kosice.upjs.sk.
\end{center}

\begin{abstract}
Using an exact mapping transformation method, magnetic properties
of a spin-1/2 and spin-1
doubly decorated Ising-Heisenberg model are investigated in detail.
By assuming that both  interaction parameters are ferromagnetic,
we found in addition to an usual ferromagnetic phase another
two phases with a nontrivial long-range order.
\vspace{1pc}
\end{abstract}

Low-dimensional quantum spin systems have become during the last few
years the subject of immense interest, since their properties
are strongly affected by quantum fluctuations. However,
the role of quantum fluctuations can be understood
from a quantum Heisenberg model (QHM) - basic theoretical model
of magnetic insulators. Although much effort has
been focused on this topic, a general exact solution has not been
found for any 2D QHM, yet.

Owing to this fact, we will propose a simplified doubly decorated
Ising-Heisenberg model, which can be exactly treated by means
of a generalized mapping transformation technique
\cite{Fish59,Stre02}.
Moreover, the suggested model can also be viewed as
useful model of magnetic insulators with the doubly decorated
network structure. Experimentally, such an arrangement of metal
ions has been recently reported in a novel
Rh$^{{\scriptsize 2+}}$-Co$^{{\scriptsize 3+}}$ compound \cite{Lu96}.
Hence, the considered model represents
a feasible model for possible structural derivatives of
the aforementioned bimetallic compound.

In this article, we will explore the Ising-Heisenberg model
with the doubly decorated network structure (see Fig.1), in which
the sites of original lattice are occupied by the spin-1/2
atoms (full circles) and the decorating sites by the spin-1
atoms (empty circles). We will further assume that the
nearest-neighbour interaction $J (\Delta)$  between spin-1 atoms is an
anisotropic Heisenberg interaction ($\Delta$ marks the
XXZ exchange anisotropy), while the nearest-neighbour
interaction $J_1$ between spin-1/2 and spin-1 atoms is purely an
Ising-type interaction.

Then, it is very convenient to write the total Hamiltonian as
a sum of bond Hamiltonians, namely $\hat {\cal H} = \sum_k \hat {\cal
H}_k$, where each bond Hamiltonian $\hat {\cal H}_k$ involves all
the interaction terms associated with $k$th couple of spin-1 atoms (Fig.1)
\begin{eqnarray}
\hat {\cal H}_k = \! \! \! &-& \! \! \! J \bigl [
  \Delta (\hat S_{k1}^x \hat S_{k2}^x + \hat S_{k1}^y \hat S_{k2}^y)
         + \hat S_{k1}^z \hat S_{k2}^z \bigr ] \nonumber \\
      \! \! \! &-& \! \! \! J_1 \hat S_{k1}^z \hat \mu_{k1}^z
          - J_1 \hat S_{k2}^z \hat \mu_{k2}^z.
\label{r1}
\end{eqnarray}
In above, $\hat \mu_k^z$ and $\hat S_{k}^{\alpha} (\alpha=x,y,z)$
denote well-known components of the standard spin-1/2 and spin-1
operators, respectively. Commutability between different bond
Hamiltonians $\hat {\cal H}_k$ ensures that the generalized
decoration-iteration transformation \cite{Fish59,Stre02}
maps the partition function ${\cal Z}_{IHM}$ of Ising-Heisenberg
model on the partition function ${\cal Z}_{IM}$ of simple spin-1/2
Ising model

\begin{equation}
{\cal Z}_{IHM} (\beta J , J_1, \Delta) = A^{Nq/2} {\cal Z}_{IM} (\beta R),
\label{r2}
\end{equation}
with following mapping parameters $A$ and $\beta R$
\begin{eqnarray}
       A^2 \! \! \! &=& \! \! \! [V (J_1, 0) + W(0)] [V(0, J_1) + W(J_1)], \nonumber \\
 \beta R \! \! \! &=& \! \! \! 2 \ln \Bigl[
                     \frac{V(J_1, 0) + W(0)}{V(0, J_1) + W(J_1)}
                                     \Bigr].
\label{r3}
\end{eqnarray}
In above, $N$ denotes a total number of spin-1/2 atoms, $q$ their
coordination number and the functions $V(x,y)$ and $W(x)$ are defined by
\begin{eqnarray}
 V(x,y) \! \! \! \! \! \! \! \! \! \! \! \! && \! \! \!
= 2 \exp(\beta J) \cosh(\beta x) + \nonumber \\
\! \! \! \! \! &+& \! \! \! 4 \cosh(\beta x/2)
                    \cosh(\beta \sqrt{y^2 + 4 (J \Delta)^2}/2),
\nonumber \\
W(x) \! \! \! &=& \! \! \!
\exp(- 2 \beta J/3) \sum_{n=1}^{3} \exp[- \beta z_n(x)],
\label{r4}
\end{eqnarray}
where $z_n$ label the roots of secular cubic equation
\begin{eqnarray}
z_n (x) \! \! \! &=& \! \! \!
2 P (x) \cos \Bigl [\phi(x) + 2 \pi (n-1)/3 \Bigr ],
\nonumber \\
P^2 (x) \! \! \! &=& \! \! \! (J/3)^2 + 2 (J \Delta)^2 /3 + x^2/3,
\nonumber \\
Q (x) \! \! \! &=& \! \! \! (J/3)^3 + J (J \Delta)^2/3 - J x^2/3,
 \\
\phi (x) \! \! \! &=& \! \! \! \frac13 \arctan
\Bigl[- \sqrt{P^6(x)-Q^2(x)}/Q(x) \Bigr].
\nonumber
\label{r5}
\end{eqnarray}
It is worthy to note that Eq.(\ref{r2}) constitutes the basic result of our calculation,
since it enables simple calculation of a large number of quantities.

To understand the behaviour of ferromagnetic Ising-Heisenberg
model ($J>0,J_1>0$) more deeply, we have examined in detail magnetizations,
correlations and quadrupolar momentum
\begin{eqnarray}
m_i^{\mp} \! \! \! \! \! &\equiv& \! \! \! \! \!
\frac12 \langle \hat \mu_{k1}^z \mp \hat \mu_{k2}^z \rangle,
\hspace*{0.4cm}
m_h^{\mp} \equiv \frac12
\langle \hat S_{k1}^z \mp \hat S_{k2}^z \rangle, \nonumber \\
C_{hh}^{xx} \! \! \! \! \! &\equiv&  \! \! \! \! \!
\langle \hat S_{k1}^x \hat S_{k2}^x \rangle,
\hspace*{0.7cm}
C_{hh}^{zz} \equiv
\langle \hat S_{k1}^z \hat S_{k2}^z \rangle,  \nonumber \\
C_{ii}^{zz} \! \! \! \! \! &\equiv& \! \! \! \! \!
\langle \hat \mu_{k1}^z \hat \mu_{k2}^z \rangle, \nonumber
\hspace*{0.7cm}
C_{ih}^{zz} \equiv \frac12
\langle \hat S_{k1}^z \hat \mu_{k1}^z + \hat S_{k2}^z \hat \mu_{k2}^z \rangle, \nonumber \\
&& \eta \equiv \frac12 \langle (\hat S_{k1}^z)^2 + (\hat S_{k2}^z)^2 \rangle,
\label{r6}
\end{eqnarray}
where $\langle ... \rangle$ denotes standard canonical averaging,
subscript $i (h)$ marks the spin-1/2 (spin-1) atom,
and superscript the spatial component of spin.

Ground-state phase diagram in the $J_1/J-\Delta$ plane
is shown Fig.2. As one can see, the parameter space
is separated by first-order transition lines into 3 regions:
the usual ferromagnetic phase (FP), valence-bond ferromagnetic phase
(VBF) and antiferromagnetic phase (AP). FP can be characterized
by saturated values of both sublattice magnetizations
($m_i^{+}=1/2$, $m_h^{+}=1$) and the spin order well-known from pure Ising model.

Contrary to this, in VBF one finds:
$m_i^{+}=1/2$, $m_h^{+}=1/2$, $C_{hh}^{zz} = 0$, $C_{hh}^{xx} = 1/2$,
$C_{ih}^{zz} = 1/4$, $C_{ii}^{zz} = 1/4$ and $\eta = 1/2$,
indicating that each couple of decorating spins
comprises of a spin '01' pair (one spin in '0' state,
another one in spin '1' state). Moreover, nonzero $C_{hh}^{xx}$
clearly implies an influence of quantum fluctuations. Hence,
it is reasonable to assume that both decorating
spin-1 atoms interchange their spin states and thus, they effectively act
as the spin-1/2 atoms \cite{Stre02}.

Finally, there also appears the unusual AP with a perfect antiferromagnetic
spin alignment in the spin-1/2 sublattice ($m_i^{-}=1/2$, $C_{ii}^{zz} = - 1/4$).
On the other hand, in the spin-1 sublattice we found the coexistence of
an antiferromagnetic long-range order along $z$ axis
($m_h^{-} < 1, C_{hh}^{zz} \ll 0$) together with a ferromagnetic
short-range ordering in $xy$ plane ($C_{hh}^{xx} \gg 0$).
To the best of our knowledge, the existence of such antiferromagnetic
long-range ordering has not been reported in any pure ferromagnetic system before.

{\it Acknowledgement}: This work was supported by APVT grant
                       No. 20-009902 and VEGA grant No. 1/9034/02

FIGURE CAPTIONS:
\newline
Fig.1: Typical example of the doubly decorated planar lattice
        (decorated square lattice). Full and empty circles denote
        the spin-1/2 and spin-1 atoms, respectively. The ellipse
        demarcates all interaction terms involved in
        $\hat {\cal H}_k$ (1).
\newline
Fig.2: Ground-state phase diagram in $J_1/J-\Delta$ plane.

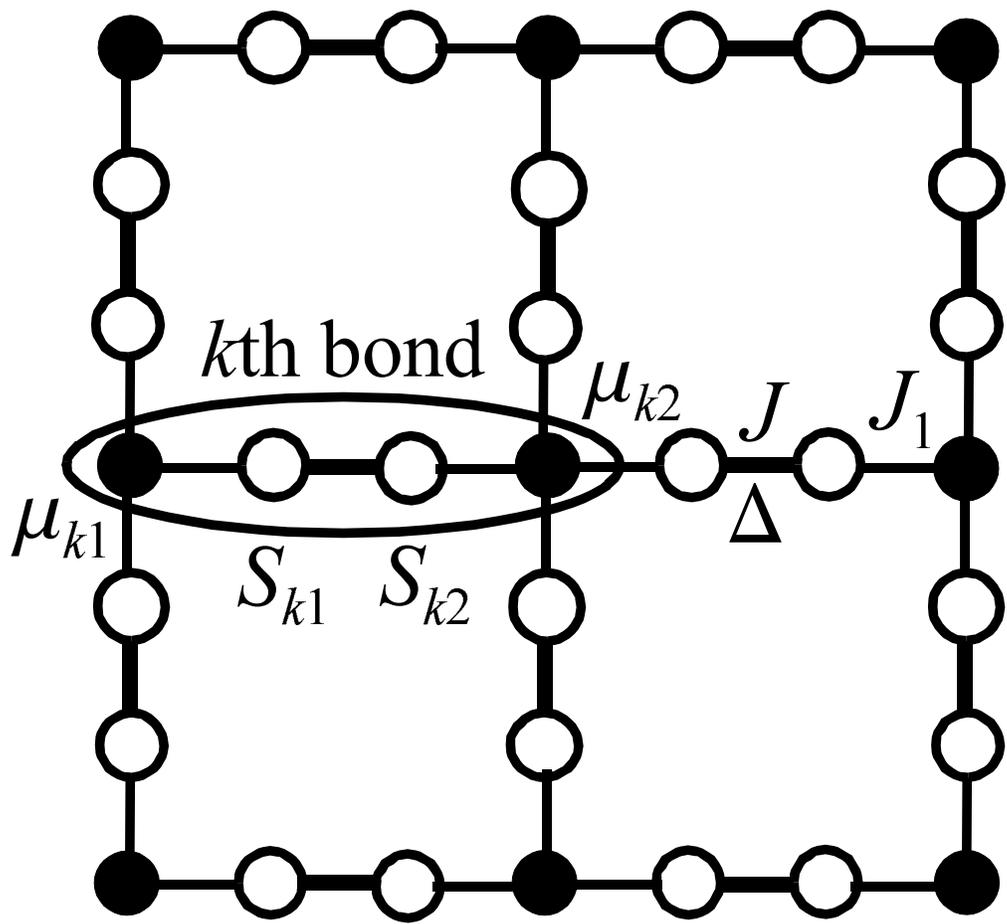
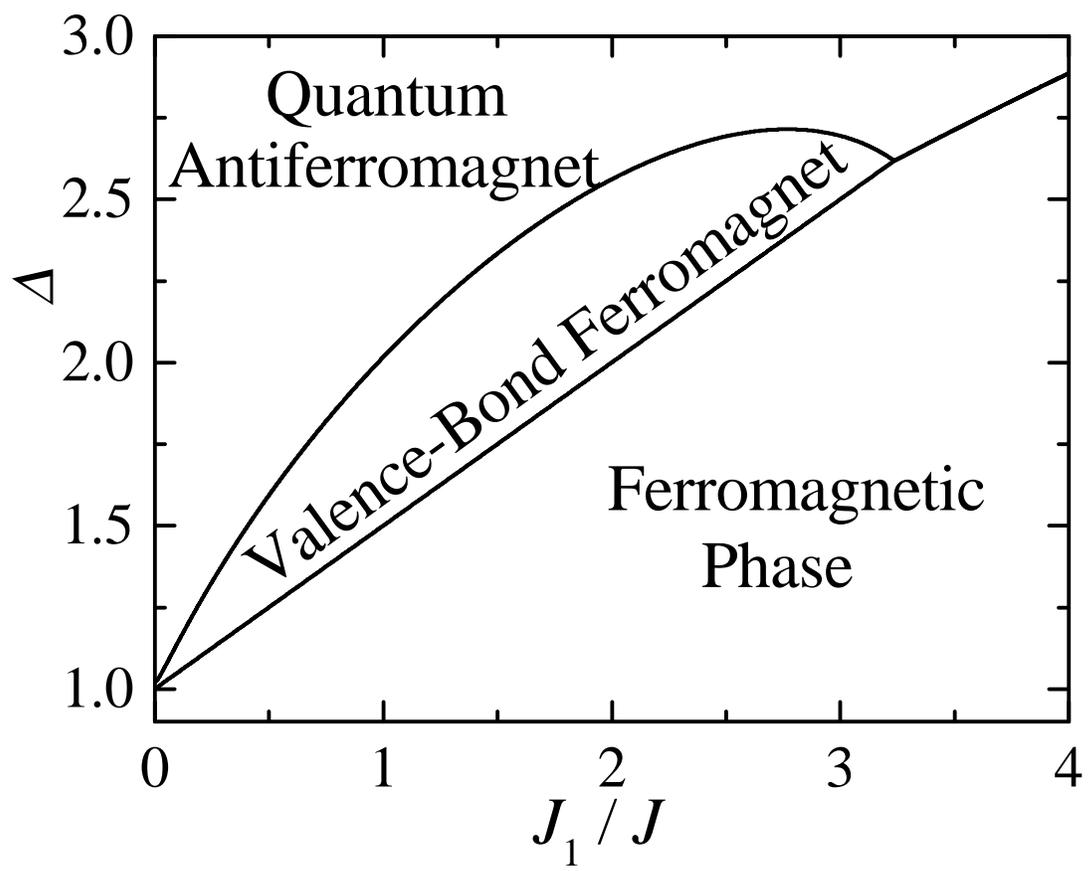
\end{document}